\documentclass[11pt,twoside]{article}
\usepackage{asp2010}

\resetcounters

\begin{document}

\title{Chromospheric Doppler Velocity Oscillations in a Sunspot}
\author{R.A.~Maurya}
\affil{Department of Physics and Astronomy, Seoul National University, Seoul 151-747, Republic of Korea.}

\begin{abstract}
We analyse the chromospheric Doppler velocity oscillations in a sunspot using the high resolution spectral observations obtained from the {\it Fast Imaging Solar Spectrograph} (FISS) of the {\it New Solar Telescope} at the Big Bear Solar Observatory. The Doppler velocity maps are constructed from the bisectors of the spectral observations. The time series analysis of  Doppler velocity maps show enhanced power in the sunspot umbra at higher frequencies and in the penumbra at lower frequencies. We find that the peak power frequency decreases gradually from the umbra to outward. 
\end{abstract}

\section{Introduction}
\label{sec:ramajor:ChromOsc:intro}

Oscillatory motions in the solar chromosphere are ubiquitous and have been studied extensively in the sunspot regions since the detection of umbral flashes \citep{Beckers1969} and running penumbral waves \citep{Zirin1972, Giovanelli1972}. The flashes are present only when the amplitude of oscillations are sufficiently strong ($\geq5\,{\rm km\,s^{-1}}$) but oscillatory motions are always present in nearly every umbra \citep{Moore1981}. The umbral flashes are believed to be due to rise in the electron density during the compressional phase of the upward propagating magneto-acoustic waves \citep{Beckers1969, Havnes1970}. The phase velocity of the magneto-acoustic wave is given by,

\begin{equation}
v=v_{\rm A}\,2^{-1/2}\left[1+r\pm\sqrt{(1+r)^2-4r^2\cos\theta_{\rm B}}\right]^{1/2}
\label{eq:ramajor:ChromOsc:PhasVel}
\end{equation}

\noindent where, $r=c^2_{s}/v^2_{A}$, $c_{\rm S}$ and $v_{\rm A}$ are the sound speed and the Alfv\'en velocity, $\theta_{\rm B}$ is the angle of wave propagation from the magnetic field direction. For a given height in the sunspot $v$ will be maximum for $\theta_{\rm B}=90^{\rm o}$, and the fast component will become magneto-sonic wave, $v_{\rm m}=\sqrt{c^2_{\rm s}+v^2_{\rm A}}$. Corresponding maximum wavelength shift \citep{Havnes1970} is given by,

\begin{equation}
\Delta\lambda_{\rm m}=\frac{\lambda_0 v_{\rm m}\cos\theta}{c} \frac{\Delta\rho_{\rm m}}{\rho_0}
\label{eq:ramajor:ChromOsc:WavlthShift}
\end{equation}

\noindent where, $\lambda_0$ is the wavelength of the undisturbed line-centre, $c$ is the speed of light, $\Delta\rho_{\rm m}/\rho_0$ is the density ratio of magneto-sonic wave, and $\theta$ is the angle between the line-of-sight and the direction of wave propagation. In case of adiabatic propagation, the temperature will change in phase with density. Rise in temperature and density caused by magneto-sonic wave will lead to increase in number density of atoms emitting the spectral line, the emission coefficient will reach maximum at line-shift (Equation~\ref{eq:ramajor:ChromOsc:WavlthShift}). Recently, it has been confirmed by \citet{Bard2010} from the NLTE simulations of Ca\,{\sc ii}\,H line.

Running penumbral waves are explained in terms of magneto-atmospheric waves \citep{Nye1974, Nye1976a}. They are fast component of the magneto-acoustic waves modified by the gravity. From recent observations, it appeared that the running penumbral waves are a sort of continuation of some of the flash waves \citep[][and references therein]{Maurya2013}. Also, we have found peak power frequency shift from umbra to outward and sudden change at umbral-penumbral boundary. However, the total time duration was shorter $\sim37$ minute giving poor frequency resolution ($\Delta\nu=0.46$\,mHz). In this article, we analyse a sunspot with longer observation period and better spatial resolution.

\section{Observational Data and Analysis}
\label{sec:ramajor:ChromOsc:DataAnaly}

We observed the active region NOAA 11553 on 27 August 2012. This active region was consists of a $\beta$-type single polarity round shaped sunspot located around at S21E06. The observations were taken using the Fast Imaging Solar Spectrograph (FISS) instrument \citep{Chae2012}, installed on a vertical optical table in the Coud\'e lab of the 1.6 meter New Solar Telescope at the Big Bear Solar Observatory, simultaneously in H$\alpha$ and Ca\,{\sc ii} 8542\AA\,(hereafter Ca\,{\sc ii}) spectral bands for a long duration of $\sim$\,6.2\,hr. A spectral data cube, with a field of view of $\sim40^{\prime\prime}$ of scan width (total number of scans 256) and slit length of $\sim40^{\prime\prime}$, covers the entire sunspot. The observations were taken under relatively good seeing conditions during 16:30:42\,–-\,22:42:53\,UT, with the scan step sampling, timing, and cadence of $\sim33$ seconds, 110 milliseconds, and $\sim36$ seconds, respectively, in both spectral bands. 

\begin{figure}[ht]
	\centering 
		\includegraphics[width=1.0\textwidth,clip=,bb=44 15 480 238]{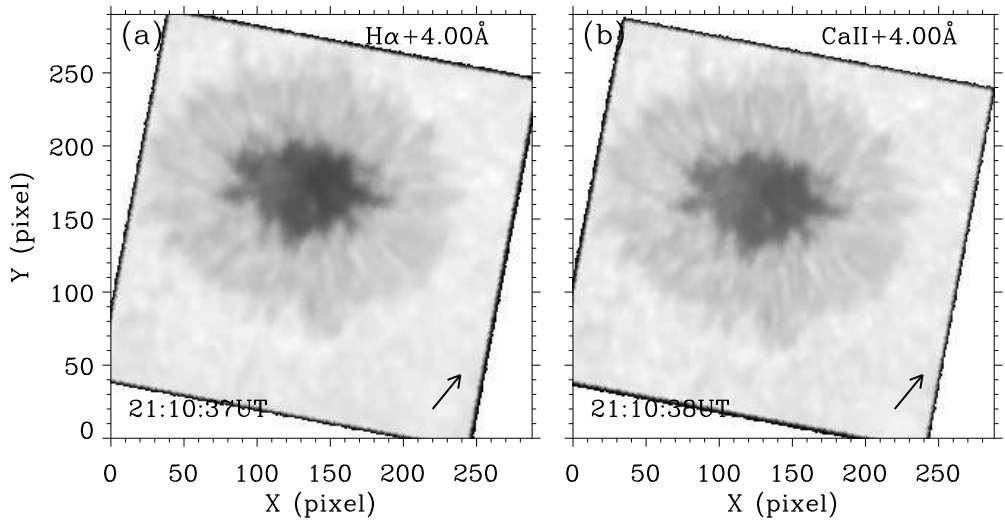}
	\caption{Intensity maps of the sunspot of the active region NOAA 11553 in \hbox{H$\alpha+4.00$\AA} (left) and Ca {\sc ii}+4.00\AA~(right) observed on 27 August 2012. Arrow marks toward the solar disk centre.}
	\label{fig:ramajor:ChromOsc:img_mosaic}
\end{figure}

All the spectral data are corrected and aligned with each other. Then chromospheric Dopplergrams are constructed using a bisector method \citep[for detail see,][]{Maurya2013}.

Figure~\ref{fig:ramajor:ChromOsc:img_mosaic} shows a sample of aligned, H$\alpha$ and Ca\,{\sc ii} intensity images at red wing wavelengths, H$\alpha$+4.0\AA~and Ca\,{\sc ii}+4.0\AA, respectively. The images have orientation from the vertical due to image rotation at Coud\'e focus. 

The image rotation causes serious problems in spectral observations: The region of interest and time-cadence changes with time, and also the time-cadence varies with position in the field-of-view \citep{Maurya2013}. Since the time cadence is not fixed, we prefer to analyse the time series using the Lomb\,--\,Scargle periodogram technique \citep{Lomb1976, Scargle1982}. For every pixel in the field of view, we computed the power spectra and constructed the power cubes for both the H$\alpha$ and Ca\,{\sc ii}.

\begin{figure}[ht]
	\centering
		\includegraphics[width=1.0\textwidth,clip=,bb=40 20 484 404]{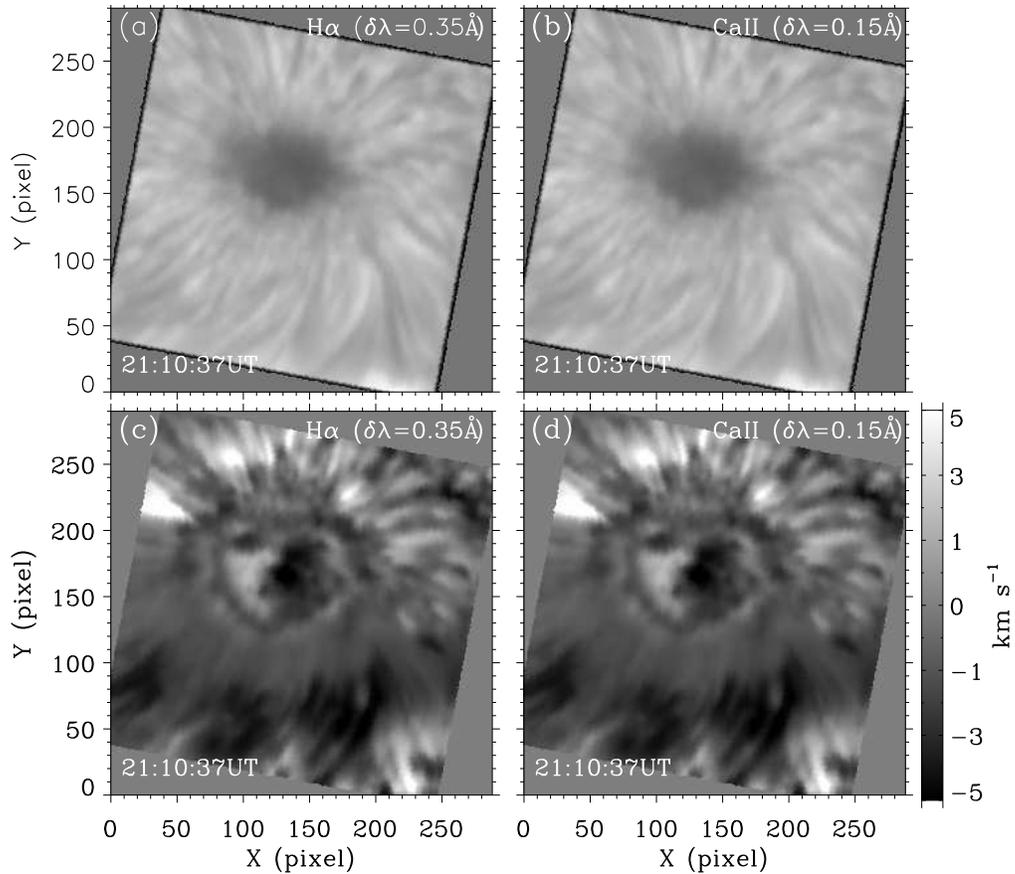}
	\caption{Chromospheric intensitygrams (top row) and Dopplergrams (bottom row) obtained from H$\alpha$ (left) and Ca\,{\sc ii} spectrograms respectively for the bisector widths ($\delta\lambda$) of 0.35\AA~and 0.15\AA.}
	\label{fig:ramajor:ChromOsc:dop_map_mosaic}
\end{figure}

\section{Results and Discussions}
\label{sec:ramajor:ChromOsc:ResDisc}

\begin{figure}[ht]
	\centering
		\includegraphics[width=1.0\textwidth,clip=,bb=44 63 518 426]{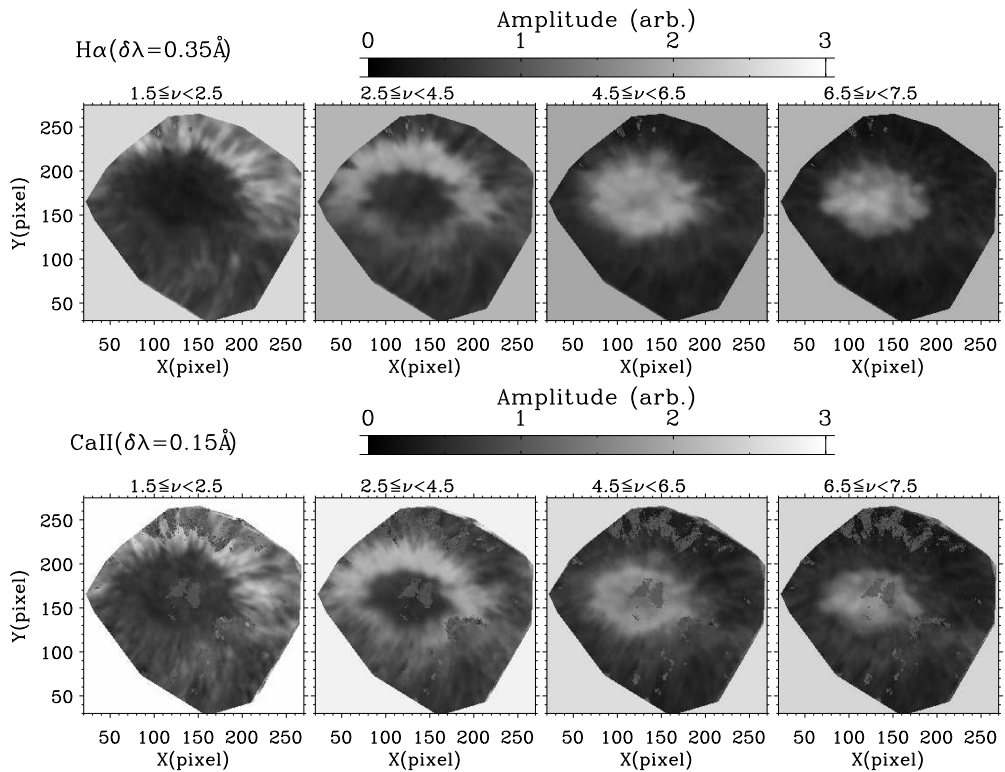}
	\caption{Average power maps in different frequency bands of the active region NOAA 11553 observed on 27 August 2012 in H$\alpha$\,(top row) and Ca\,{\sc ii}\,(bottom row)}.
	\label{fig:ramajor:ChromOsc:pw_map_frq}
\end{figure}

\begin{figure}[ht]
	\centering
		\includegraphics[width=1.0\textwidth,clip=,bb=30 3 432 230]{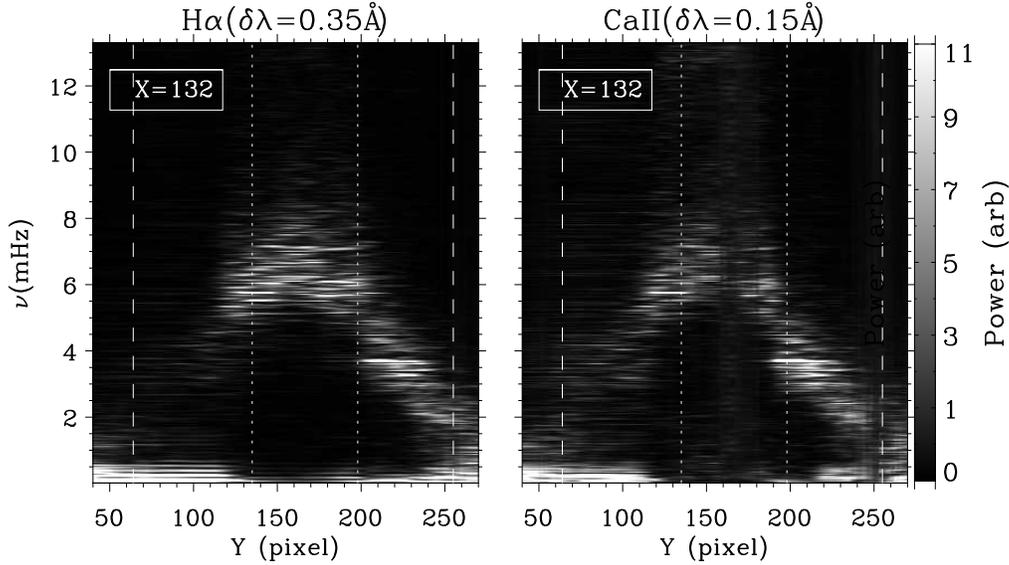}
	\caption{Power maps as function of space (along y-direction at x=132) and frequency. Vertical dotted and dashed lines mark the locations of the umbral and penumbral boundaries, respectively, as measured from intensity images (see Figure~\ref{fig:ramajor:ChromOsc:img_mosaic}).}
	\label{fig:ramajor:ChromOsc:space_line_pwmap}
\end{figure}

The results of the analysis are shown in the Figures~\ref{fig:ramajor:ChromOsc:img_mosaic}--~\ref{fig:ramajor:ChromOsc:space_line_pwmap}.

Figure~\ref{fig:ramajor:ChromOsc:dop_map_mosaic} shows samples of chromospheric Dopplergrams and intensitygrams for H$\alpha$ and Ca\,{\sc ii} at bisector widths, $\delta\lambda=0.35$\AA~and 0.15\AA, respectively. The ring-like pattern in the penumbra in Dopplergrams shows the running penumbral waves. These are not clearly seen in the intensity maps. In the sunspot umbra there are high velocity patterns which show umbral flashes. There is an asymmetry in the Doppler velocities of the upper and lower part of the sunspot caused by off-disk-centre viewing. The negative Doppler velocity in the limbward part of the penumbra indicates the line-of-sight component of the reverse Evershed flow. 

Figure~\ref{fig:ramajor:ChromOsc:pw_map_frq} shows sample of power maps averaged in different frequency bands. At lower frequencies there is enhanced power around the sunspot umbra. As we move from lower to higher frequency bands the peak power shift towards umbra. At higher frequency band ($6.5\leq\nu<7.5$\,mHz) all the power are concentrated in the umbral location. These features are also seen in the Ca\,{\sc ii} power maps except some noisy structures in the umbra and penumbra. These noisy patches are caused by emission features, such as umbral flashes, which lead to failure of bisector method for Doppler velocity computation. 

Figure~\ref{fig:ramajor:ChromOsc:space_line_pwmap} shows the power maps as function of space (along y-direction at x=133) and frequency. These maps show that the peak power frequency decreases from umbra to outward. Similar, maps are also shown in \citet{Maurya2013} but for the half side of a sunspot, also there is no boundary or gap between 3- and 5- minute bands as found in previous study. This may be due to better frequency resolution ($\Delta\nu\approx0.045$\,mHz); but in our previous study, $\Delta\nu\approx0.46$\,mHz. The linear trend of the peak power frequency shift, in both the spectral bands, is a very interesting pattern.  

In the sunspot, we have MHD waves with all the restoring forces, Lorentz force, plasma pressure, and gravity. Furthermore, the chromospheric sunspot umbra and penumbra may not form at same geometrical height. Thus, the frequency shift from umbra to outward may be due to different physical conditions. One of the possible explanation for the frequency variation, from the umbra to outward, can be given based on the cutoff frequency of the MHD waves in sunspot. 

For upward propagating wave (along the magnetic field lines), the cutoff frequency is given by,

\begin{equation}
\nu_{\rm cutoff}=\frac{\gamma\,g}{2\,c_{\rm s}}
\label{eq:ramajor:ChromOsc:CutoffFreq}
\end{equation}

\noindent where, $g$ is the acceleration due to gravity, and $\gamma$ is the ratio of specific heats. $\nu_{\rm cutoff}$ will not be affected by the magnetic field because the plasma oscillations will be parallel to it where Lorentz force will not work. This condition will meet in the sunspot umbra where the field lines are almost vertical. 

For propagation at any angle from the vertical ($\theta_{\rm g}$), the Lorentz force will play an important role. The slow mode will propagate parallel to the field lines where they will experience the vertical component of $g$, i.e., $g\cos\theta_{\rm g}$, hence, $\nu_{\rm cutoff}\propto\cos\theta_{\rm g}$. This shows that the increase in inclination (in range $0<\theta_{\rm g}<90$) will lead to decrease in $\nu_{\rm cutoff}$, i.e., if we consider only the effects of gravity, the cutoff frequency will decrease from the umbra to outward  because of increase in field inclination. This explains the peak power frequency pattern as seen in Figure~\ref{fig:ramajor:ChromOsc:space_line_pwmap}. 

\section{Summary and Conclusions}
\label{sec:ramajor:ChromOsc:SumConc}

We analysed chromospheric Doppler velocity time series computed from the bisectors of spectral observations in the H$\alpha$ and Ca\,{\sc ii}\,8542\AA. We found enhanced power in the sunspot umbra at higher frequencies and in the penumbra at lower frequencies. The peak power frequency decreases gradually from umbra to outward which confirms our previous reports \citep{Maurya2013}. But we do not find frequency shift at umbral-penumbral boundary from three to five minutes bands as reported previously. The decrease in peak power frequency from umbra to outward is an interesting pattern. We plan to analyse several sunspots for frequency changes, the Doppler velocity asymmetry, frequency resolved phases, umbral flashes, etc. It would be interesting to analyse the variation in sunspot oscillation associated with activities, and in different features of the chromosphere.
 
\acknowledgements 
 This work was supported by the National  Research Foundation of Korea (2011-0028102).


\begin{thebibliography}{}
\expandafter\ifx\csname natexlab\endcsname\relax\def\natexlab#1{#1}\fi
\expandafter\ifx\csname url\endcsname\relax
  \def\url#1{\texttt{#1}}\fi
\expandafter\ifx\csname urlprefix\endcsname\relax\def\urlprefix{URL }\fi
\providecommand{\eprint}[2][]{\url{#2}}

\bibitem[{{Bard} \& {Carlsson}(2010)}]{Bard2010}
{Bard}, S., \& {Carlsson}, M. 2010, \apj, 722, 888

\bibitem[{{Beckers} \& {Tallant}(1969)}]{Beckers1969}
{Beckers}, J.~M., \& {Tallant}, P.~E. 1969, \solphys, 7, 351

\bibitem[{{Chae} et~al.(2012){Chae}, {Park}, {Ahn}, {Yang}, {Park}, {Nah},
  {Jang}, {Cho}, {Cao}, \& {Goode}}]{Chae2012}
{Chae}, J., {Park}, H.~M., {Ahn}, K., {Yang}, H., {Park}, Y.~D., {Nah}, J.,
  {Jang}, B.~H., {Cho}, K.~S., {Cao}, W., \& {Goode}, P.~R. 2012, \solphys,
  published online

\bibitem[{{Giovanelli}(1972)}]{Giovanelli1972}
{Giovanelli}, R.~G. 1972, \solphys, 27, 71

\bibitem[{{Havnes}(1970)}]{Havnes1970}
{Havnes}, O. 1970, \solphys, 13, 323

\bibitem[{{Lomb}(1976)}]{Lomb1976}
{Lomb}, N.~R. 1976, \apss, 39, 447

\bibitem[{Maurya et~al.(2013)Maurya, Chae, Park, Yang, Song, \&
  Cho}]{Maurya2013}
Maurya, R.~A., Chae, J., Park, H., Yang, H., Song, D., \& Cho, K. 2013,
  \solphys, online

\bibitem[{{Moore}(1981)}]{Moore1981}
{Moore}, R.~L. 1981, in The Physics of Sunspots, edited by L.~E. {Cram}, \&
  J.~H. {Thomas}, 259

\bibitem[{{Nye} \& {Thomas}(1974)}]{Nye1974}
{Nye}, A.~H., \& {Thomas}, J.~H. 1974, \solphys, 38, 399

\bibitem[{{Nye} \& {Thomas}(1976)}]{Nye1976a}
--- 1976, \apj, 204, 582

\bibitem[{{Scargle}(1982)}]{Scargle1982}
{Scargle}, J.~D. 1982, \apj, 263, 835

\bibitem[{{Zirin} \& {Stein}(1972)}]{Zirin1972}
{Zirin}, H., \& {Stein}, A. 1972, \apjl, 178, L85

\end{thebibliography}
\end{document}